\documentclass[aps,twocolumn,a4paper,showpacs,superscriptaddress]{revtex4}
\usepackage{graphicx}
\usepackage{amssymb}
\usepackage{amsmath}
\newcommand{\be}{\begin{equation}}
\newcommand{\ee}{\end{equation}}
\newcommand{\ben}{\begin{eqnarray}}
\newcommand{\een}{\end{eqnarray}}
\newcommand{\bes}{\begin{subequations}}
\newcommand{\ees}{\end{subequations}}
\newcommand{\bb}{\bibitem}

\newcommand{\bfi}{\begin{figure}}
\newcommand{\efi}{\end{figure}}
\newcommand{\bc}{\begin{center}}
\newcommand{\ec}{\end{center}}
\newcommand{\xx}{\sqrt{1-\frac{2X}{M^2}}}

\newcommand{\LL}{{\cal L}}
\begin{document}
\title{Twinlike Models in Scalar Field Theories}
\author{D. Bazeia}
\affiliation{Departamento de F\'{\i}sica, Universidade Federal
da Para\'{\i}ba, 58051-970 Jo\~ao Pessoa, PB, Brazil}
\author{J. D. Dantas}
\affiliation{Departamento de F\'{\i}sica, Universidade Federal
da Para\'{\i}ba, 58051-970 Jo\~ao Pessoa, PB, Brazil}
\affiliation{Unidade Acad\^emica de Educa\c c\~ao, Universidade Federal de Campina Grande, 58175-000 Cuit\'e, PB, Brazil}
\author{A. R. Gomes}
\affiliation{Departamento de Ci\^encias Exatas, Centro Federal de Educa\c c\~ao Tecnol\'ogica do Maranh\~ao, 65030-000 S\~ao Lu\'{\i}s, MA, Brazil}

\author{L. Losano}
\affiliation{Departamento de F\'{\i}sica, Universidade Federal
da Para\'{\i}ba, 58051-970 Jo\~ao Pessoa, PB, Brazil}

\author{R. Menezes}
\affiliation{Departamento de Ci\^encias Exatas, Universidade Federal
da Para\'{\i}ba, 58297-000 Rio Tinto, PB, Brazil}
\affiliation{Departamento de F\'{\i}sica, Universidade Federal de Campina Grande, 58109-970, Campina Grande, Para\'\i ba, Brazil}

\date{\today}

\begin{abstract}
This work deals with the presence of defect structures in models described by real scalar field in a diversity of scenarios. The defect structures which we consider are static solutions of the equations of motion which depend on a single spatial dimension. We search for different models, which support the same defect solution, with the very same energy density.
We work in flat spacetime, where we introduce and investigate a new class of models. We also work in curved spacetime, within the braneworld context, with a single extra dimension of infinite extent, and there we show how the brane is formed from the static field configuration.
\end{abstract}

\pacs{11.27.+d, 11.10.Kk}

\maketitle
\section{Introduction}

Topological structures appear in several areas of nonlinear science. In high energy physics, the most known topological structures are kinks, vortices and monopoles.
Kinks are the simplest structures, and they appear in one spatial dimension, in the presence of a single real scalar field. Vortices and monopoles are more involved, and they usually appear in two and three spatial dimensions, in the presence of complex scalar fields coupled to Abelian and non Abelian gauge fields, respectively \cite{book1,book2}. 

In the present work we focus our attention on kinks in relativistic systems described by real scalar field. The subject has been investigated in a diversity of contexts, but here we bring novelties related to the issue recently studied by  Andrews, Lewandowski, Trodden and Wesley (ALTW) in Ref.~{\cite{altw}}, concerning the presence of twin models, that is, of distinct models which support the very same topological structure. To make the investigation more general, however, we will consider models engendering topological or kinklike and nontopological or lumplike structures.

Relativistic models described by scalar fields have been studied in a diversity of scenarios. Here we quote some works dealing with junctions of kinklike structures \cite{gt,sakai,trodden1,bb1,bb2}, the deformation procedure which lead us to new models and the respective kinklike solutions \cite{dd,ddd}, the presence of vortons \cite{shellard} and other new models and solutions \cite{bmm,gleiser1,gleiser2,dunne,usal,vacha}. Also, from the experimental point of view, new and interesting kinklike solutions have been found recently in systems constrained to work within small regions, at the nanometer scale \cite{exp1,exp2}.

Motivated by cosmology, another direction of study concerns models where kinematic modifications of the scalar field are introduced, attempting to contribute to explain the present accelerated expansion of the Universe \cite{c1,c2,c3,c4}. Recent investigations of scalar fields with modified kinematics have been presented in \cite{s1,s2,s3,b1,b2,b3,b4} 
with distinct motivations, in particular to search for the presence of compacton, that is, of topological structures which live in a compact region, and for applications in the braneworld context, within the five dimensional spacetime engendering a single extra dimension of infinite extent \cite{RS,GW,F}.

In the study of models with modified kinematics, in \cite{altw} the authors noted that it is possible for k-defects to masquerade as canonical
scalar field solutions. That is, given a standard scalar field model described by the field $\phi$, with canonical kinetic term and potential $V(\phi)$ in the usual form,
the topological defect profile $\phi(x)$ and corresponding energy density $\rho(x)$ can also appear in a k-defect theory. We call the modified or k-defect theory introduced in \cite{altw}, the ALTW model. And we recall that, despite having identical defect solutions, the two theories are not reparametrizations of each other, since the fluctuation spectra about the kink are different. In the current work, we call the two distinct models twinlike theories, because they support the very same defect structures, engendering the very same energy density.

In this work we further investigate this issue, considering as twinlike theories the standard model and the ALTW model introduced in \cite{altw}. We deal with the defect solutions, and we compare the two models in the case where the static solutions are more general. We do this studying solutions of the equations of motion in the $(\phi,\phi^\prime)$ plane, where prime stands for spatial derivative of the field, $\phi^\prime=d\phi/dx$, with $\phi(x)$ describing static solution. Also, we extend the results to the case of non-topological or bell-shape solutions, which are of direct interest to tachyon condensation in string theory \cite{sen1,z1,z2,sen2}, and to cosmology \cite{sen3}. 

We take advantage of the formalism introduced in Ref.~\cite{b3}, from which it is possible to get to first-order differential equation that solves the equation of motion for defect solutions of a general scalar field model. We use this to verify if the issue is specific of the two models introduced in \cite{altw}. In the process, we further ask for other models, with the answer offering a new class of twinlike models. We also study the issue in curved spacetime, coupling the scalar field with gravity within the braneworld context, with a single extra dimension of spatial nature and infinite extent.

To ease understanding of the subject, we organize the work as follows: In the next Sec.~\ref{sec2} we explore the standard and the ALTW models, and we comment on some specific properties the two models engender. In Sec.~\ref{sec3} we investigate two distinct ways to distinguish twinlike models. In Sec.~\ref{sec4} we generalize the investigation to another class of twinlike models, and in Sec.~\ref{sec5} we deal with twinlike models in the braneworld context. We end the work in Sec.~\ref{sec6}, where we include comments and conclusions.

\section{The models}\label{sec2}

Let us now focus attention on the standard and modified models. We first consider the standard model, and then we deal with the modified model. We also comment on some specific properties the two models engender. 

\subsection{Standard model}

The standard model is described by the Lagrange density
\ben\label{standard}
{\cal L}_s &=&\frac12\partial_\mu\phi\partial^\mu\phi-V(\phi),\nonumber\\
&=&X-V(\phi),
\een
where
\be
X=\frac12\partial_\mu\phi\partial^\mu\phi,
\ee
and $V(\phi)$ is the potential. The equation of motion has the form
\be
\partial_\mu\partial^\mu\phi+V_\phi=0,
\ee
where $V_\phi$ stands for the derivative of the potential with respect to the field $\phi$, that is, $V_\phi=dV/d\phi$.
The energy-momentum tensor has the form
\be
T^{\mu\nu}=\partial^\mu\phi\partial^\nu\phi-\eta^{\mu\nu}{\cal L}_s.
\ee
Here we work in $(1,1)$ space-time dimensions, with the metric obeying ${\rm diag}\,\eta^{\mu\nu}=(1,-1)$. The equation of motion for static solution becomes
\be
\phi^{\prime\prime}=V_\phi,
\ee
where $\phi^\prime=d\phi/dx$, etc. And the energy-momentum tensor reduces to
\be
T^{00}=\rho(x)=\frac12\phi^{\prime\,2 }+ V(\phi),
\ee
which is the energy density, and
\be
Tö^{11}=\frac12\phi^{\prime\,2 }- V(\phi).
\ee

As an example, we consider the potential with quartic self-interaction, in the form
\be\label{v4}
V_4(\phi)=\frac{1}{2}\left(1-\phi^2\right)^2,
\ee
taking the field and coordinates as dimensionless quantities. 
This model is well known, and it has the minima
$\bar{\phi}_{\pm}=\pm 1$. Also, it supports the defect solutions
\be
\phi_{\pm}(x)=\pm\tanh(x-\bar{x}),
\ee
where $\bar{x}$ stands for the center of the defect, which we take to be at the origin in the $x$-axis, that is, $\bar{x}=0$. The two minima $\bar{\phi}_{\pm}$
define a topological sector, and the above solutions are topological or kinklike defects.

We can consider another model, with potential
\be
V_{sG}(\phi)=\frac12\cos^2(\phi).
\ee 
This is the sine-Gordon model, and typical solutions are
\be
\phi(x)=\pm \arcsin(\tanh(x))+k\pi, \,\,\,\, k=0, \pm1, \pm2, \ldots
\ee

We can use another potential, with cubic self-interaction, in the form
\be\label{pot3}
V_3(\phi)=\frac12 \phi^2\left( 1- \phi\right).
\ee 
This model has a single, local minimum, at $\bar {\phi}=0$, and it supports the defect solution
\be
\phi(x)={\rm sech}^2(x).
\ee
This solution describes a nontopological or lumplike defect. Note that \eqref{pot3} has a partner potential, with the factor $(1-\phi)$
changed to $(1+\phi)$; the partner model supports the non-topological defect $\phi(x)=-{\rm sech}^2(x)$.

Another model is described by the potential
\be
V_{4i}(\phi)=\frac12\phi^2(1-\phi^2).
\ee
It is an inverted $\phi^4$ model, and it has a local minimum at ${\bar \phi}=0$; it supports the nontopological or lumplike defects
\be
\phi(x)=\pm{\rm sech}(x).
\ee

\subsection{The ALTW model}

Let us now turn our attention to the modified model which was considered in Ref.~\cite{altw}. It is described by the Lagrange density
\ben\label{altw}
{\cal L}_m=M^2-M^2\left(1+\frac{U(\phi)}{M^2}\right)\xx.
\een
We first note that in the limit where the mass scale $M^2$ is very large, we get from the modified Lagrange density, up to the order $1/M^2$:
\be
\LL_m\approx X - U(\phi)+\frac{1}{M^2}\left(\frac12 X^2 +U(\phi)X\right).
\ee 
Thus, we see that the limit $1/M^{2}\to0$, leads us back to the standard model \eqref{standard}, with the identification $U(\phi)=V(\phi)$, so we call $U(\phi)$
the potential of the modified model.

In general, for the modified model \eqref{altw}, the equation of motion is given by
\be
\partial_\mu \left(\left(\frac{1+\frac{U(\phi)}{M^2}}{\xx}\right)\partial^\mu \phi\right)+{U_\phi} \xx=0,
\ee
or more explicitly
\be\label{eqofmotion1}
\left(\eta^{\mu\nu} + \frac{1}{M^2}\frac{\partial^\mu \phi \partial^\nu \phi}{1-\frac{2X}{M^2}}\right)
\partial_\mu \partial_\nu \phi = - \frac{1}{M^2}\frac{U_\phi}{1+\frac{U(\phi)}{M^2}}.
\ee 
Here the energy-momentum tensor has the form
\ben\label{Tmunu}
T^{\mu\nu}\!&=&\!\frac{(1\!+\!\frac{U(\phi)}{M^2})\partial^\mu \phi \partial^\nu \phi}{\xx}\nonumber\\
&&- M^2\eta^{\mu\nu} \!\left(\!1\!-\!\left(\!1\!+\!\frac{U(\phi)}{M^2}\right)\!\xx\right)\! .
\een

We focus our attention on static configurations, with $\phi=\phi(x)$; in this case, the equation of motion becomes
\be\label{staticeq1}
\frac{\phi^{\prime\prime}}{1+\frac{\phi^{\prime2}}{M^2}}=-\frac{1}{M^2}\frac{U_\phi}{1+\frac{U(\phi)}{M^2}}.
\ee
We integrate this equation once, to get the first-order equation
\be\label{stressless}
1-\frac{1+\frac{U(\phi)}{M^2}}{\sqrt{1+\frac{\phi^{\prime2}}{M^2}}}=\frac{C}{M^2},
\ee
where $C$ is an integration constant. This equation can be rewritten as  
\be\label{genra}
\phi^{\prime2}=-M^2 + M^2\left(\frac{1+\frac{U(\phi)}{M^2}}{1-\frac{C}{M^2}}\right)^2.
\ee

Since the energy-momentum tensor is conserved, for static field we get that  $T^{11}$ has to be a constant, which leads us to Eq.~\eqref{stressless}.
On the other hand, the energy density $T^{00}=\rho(x)$ has the form
\bes\label{ddG}
\be\label{firstorder1}
\rho(x)=-M^2+M^2\left(1+\frac{U(\phi)}{M^2}\right)\sqrt{1+\frac{\phi^{\prime2}}{M^2}},
\ee
or
\ben\label{rho1}
\rho(x)&=&-M^2+M^2\frac{\left(1+\frac{U(\phi)}{M^2}\right)^2}{1-\frac{C}{M^2}},\nonumber \\
&=&-M^2+M^2\left(1-\frac{C}{M^2}\right)\left(1+\frac{\phi^{\prime2}}{M^2}\right).
\een
\ees
In Ref.~\cite{s2} one shows that a necessary condition to have stable kinklike solutions is to set $C=0$. In this case, the equation \eqref{genra} becomes
\be
\phi^{\prime2}=2U(\phi)+\frac{U^2(\phi)}{M^2},
\ee
and the energy density gets to the form
\be\label{rhoT}
\rho(x)=2U(\phi)+\frac{U^2(\phi)}{M^2}=\phi^{\prime2}.
\ee

\subsection{Some specific properties}

Let us now follow \cite{altw} and consider the standard model, with the specific potential
\be\label{new}
V(\phi)=U(\phi)+\frac{1}{2}\frac{U^2(\phi)}{M^2},
\ee
where $U(\phi)$ is the potential for the modified model. We note in the standard model with the above potential, the equation of motion is given by
\be
\eta^{\mu\nu} \partial_\mu \partial_\nu \phi +\left(1+\frac{U(\phi)}{M^2}\right)U_\phi=0,
\ee
and the energy-momentum tensor has the form
\be\label{Tnew}
T^{\mu\nu}=\partial^\mu \phi \partial^\nu \phi - \eta^{\mu\nu}X+\eta^{\mu\nu}\left(U(\phi)+\frac12 \frac{U^2(\phi)}{M^2}\right).
\ee
In the case of a static solution, we get
\be\label{eom}
\phi^{\prime\prime} =\left(1+\frac{U(\phi)}{M^2}\right)U_\phi,
\ee
and
\be\label{ene}
\rho(x)=\frac12 \phi^{\prime 2}+U(\phi)+\frac12 \frac{U^2(\phi)}{M^2}.
\ee
However, from \eqref{Tnew} we can write 
\be
T^{11}=\frac12\phi^{\prime 2}-U(\phi)-\frac12\frac{U^2(\phi)}{M^2}.
\ee
Since the energy-momentum is conserved, for static solution we can write $T^{11}=C$, and this leads us to the equation
\be\label{C1eqref}
\phi^{\prime 2}=2C+2 U(\phi)+ \frac{U^2(\phi)}{M^2},
\ee
and  for $C=0$ one gets to
\be\label{foeq}
\phi^{\prime 2}=2U(\phi)+\frac{U^2(\phi)}{M^2}.
\ee
We see that solutions of the above first-order equation solve the equation of motion \eqref{eom}.
Thus, we reproduce the result of Ref.~{\cite{altw}}, showing that the modified model \eqref{altw} and the model \eqref{new} give the same defect structure.
Moreover, using \eqref{foeq} in \eqref{ene} we reproduce \eqref{rhoT}, and so the defect has the very same energy density, as also stated in Ref.~\cite{altw}.

The above result requires that we make $T^{11}=0$. In the more general case, we can use $T^{11}=C$, and this makes the two models different. Therefore, a nice way
to distinguish between the two theories is simply to investigate the solutions taking $T^{11}=C$, as a constant but nonvanishing quantity. We explore this fact in the next Sec.~\ref{sec4}.

We now take advantage of the result firstly presented in Ref.~\cite{b3}, from which one can write a first-order framework for defect solutions of a general scalar field model. The approach concerns introducing the function $W(\phi)$, such that
\be
\LL_X \phi^\prime=W_\phi,
\ee
for some generic Lagrange density ${\cal L}$. In the case of the standard Lagrange density, we have ${\cal L}_{s X}=1$, and this leads to
the equation
\be
\phi^\prime=W_\phi,
\ee
such that
\be
V(\phi)=\frac12 W_\phi^2.
\ee
This is the first-order framework one usually gets for scalar field with standard dynamics.

However, the very same framework can be used to study the modified model. In this case, taking $\LL_{m X} \phi^\prime=W_\phi$ leads to the equation
\be
\phi^{\prime 2}=\frac{W^2_\phi}{\left(1+\frac{U(\phi)}{M^2}\right)^2-\frac{W_\phi^2}{M^2}}.
\ee
To make this equation compatible with \eqref{genra} for $C=0$, we have to write
\be\label{upot}
U(\phi)=-M^2+M^2\sqrt{1+\frac{W^2_\phi}{M^2}}.
\ee
An important point in the above procedure is that the energy density, which can be written as $\rho(x)=\phi^{\prime 2}$, leads us to the nice result
\be
\rho(x)=\phi^{\prime 2}=W_\phi\,\phi^\prime=\frac{dW}{dx}.
\ee
It is a total derivative, so the energy follows easily, in the form $E=| \Delta W|=|W(\phi(\infty)-W(\phi(-\infty))|$. 

A nice information follows when one expands \eqref{upot} for small $1/M^2$; up to the lowest order we get
\be
U(\phi)=\frac12 W_\phi^2.
\ee
On the other hand, from \eqref{new}, the limit $1/M^2\to0$ leads us to $V(\phi)=U(\phi)$. Thus, we have the same $W=W(\phi)$ controlling both the standard and modified models, although $V(\phi)$ and $U(\phi)$ are different functions of the real scalar field.

\section{Distinguishing Twin Theories}
\label{sec3}

Let us now focus on the issue of distinguishing the standard and modified models. In \cite{altw}, the authors showed that the study of linear stability of the defect structure
is able to distinguish the two models. Since linear stability is important for introducing quantum corrections at the one-loop level, we can say that the quantum effects distinguish the two models. More than that, we have already shown that there is another way to distinguish the two models, if one deals with a constant but non vanishing $T^{11}\neq0$.
Thus, below we investigate the two possibilities, of using a constant but nozero $T^{11}$, and of studying linear stability. However, since linear stability was already investigated in \cite{altw}, here we depart from \cite{altw} and follow an alternative way to investigate linear stability.

\subsection{Nonzero $T^{11}$}

Let us firstly consider the case of a non vanishing $T^{11}$. In this case, the equation of motion for the standard and modified models are, respectively,
\bes
\ben\label{eq1}
\phi^{\prime 2}&=&W_\phi^2+2C,\\
\phi^{\prime2}&=&-M^2+M^2\frac{1+\frac{W_\phi^2}{M^2}}{\left(1-\frac{C}{M^2}\right)^2}.\label{eq2}
\een
\ees
We take as an explicit example, the model described by
\be
W_\phi=1-\phi^2.
\ee
Note that for the standard model, the potential $V(\phi)$ leads to the $\phi^4$ model, with spontaneous symmetry breaking; see Eq.~\eqref{v4}. With this choice, the first order equations \eqref{eq1} and \eqref{eq2} become
\bes\label{phasespace}
\ben
\phi^{\prime2}&=&(1-\phi^2)^2 + 2C,\\
\phi^{\prime2}&=&-1 + \frac{1+\frac{(1-\phi^2)^2}{M^2}}{\left(1-\frac{C}{M^2}\right)^2},
\een\ees
for the standard and modified models, respectively.  We plot in Fig.~1 the solutions in the $(\phi,\phi^\prime)$ plane, for $C>0$, $C=0$, and $C<0$, with dashed (red), solid (black) and dotted (blue) lines, respectively. Here we note that only the black curves, which correspond to kink and antikink, are identical in both plots. We note that the two phase diagrams tend to become the same as $1/M^2$ becomes greater and greater. The solid (black) curves correspond to the topological kink and antikink configurations
\be
\phi(x)=\pm \tanh(x).
\ee 

\begin{figure}[t]
\centering
\includegraphics[width=3.2cm]{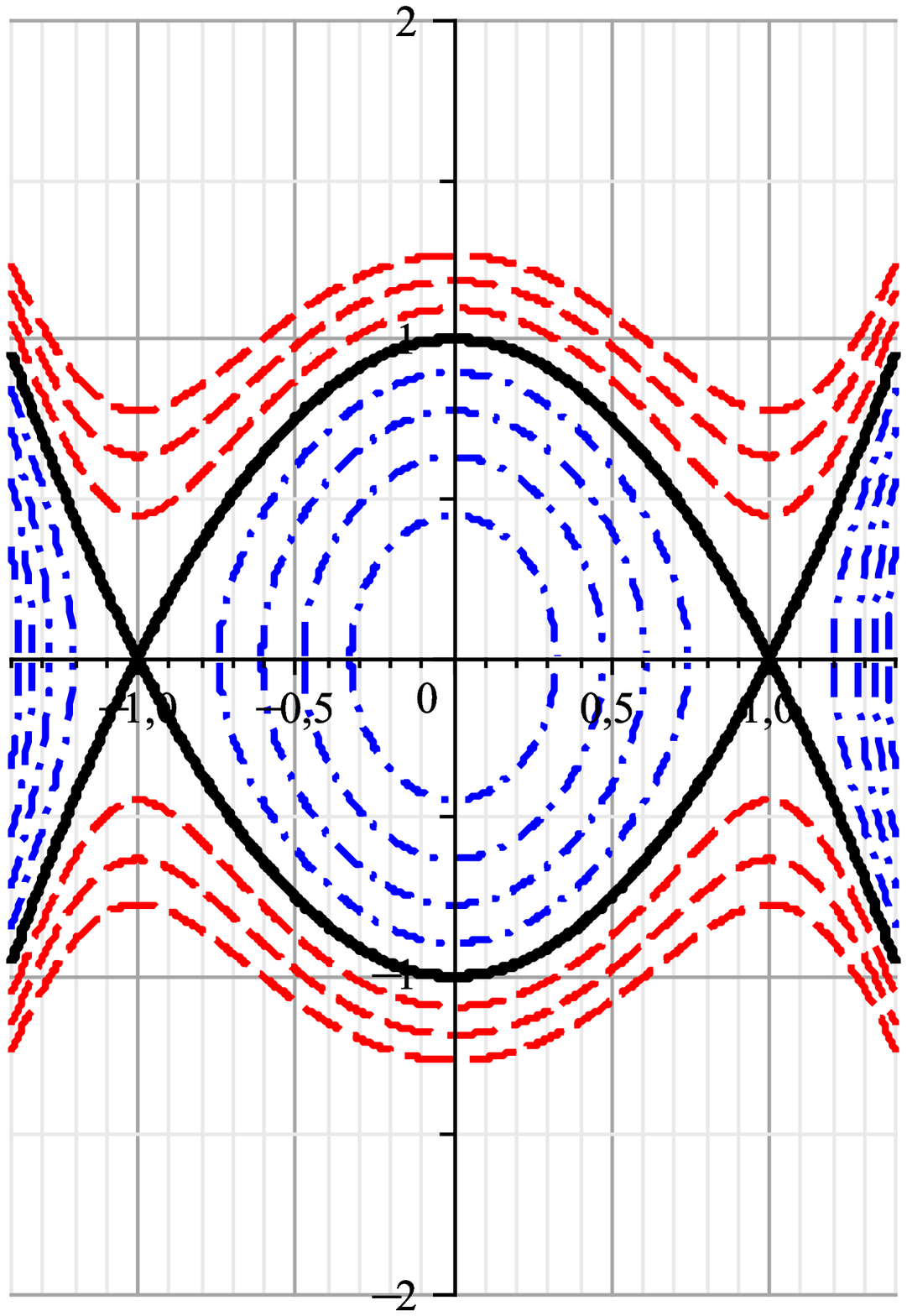}\hspace{0.8cm}
\includegraphics[width=3.2cm]{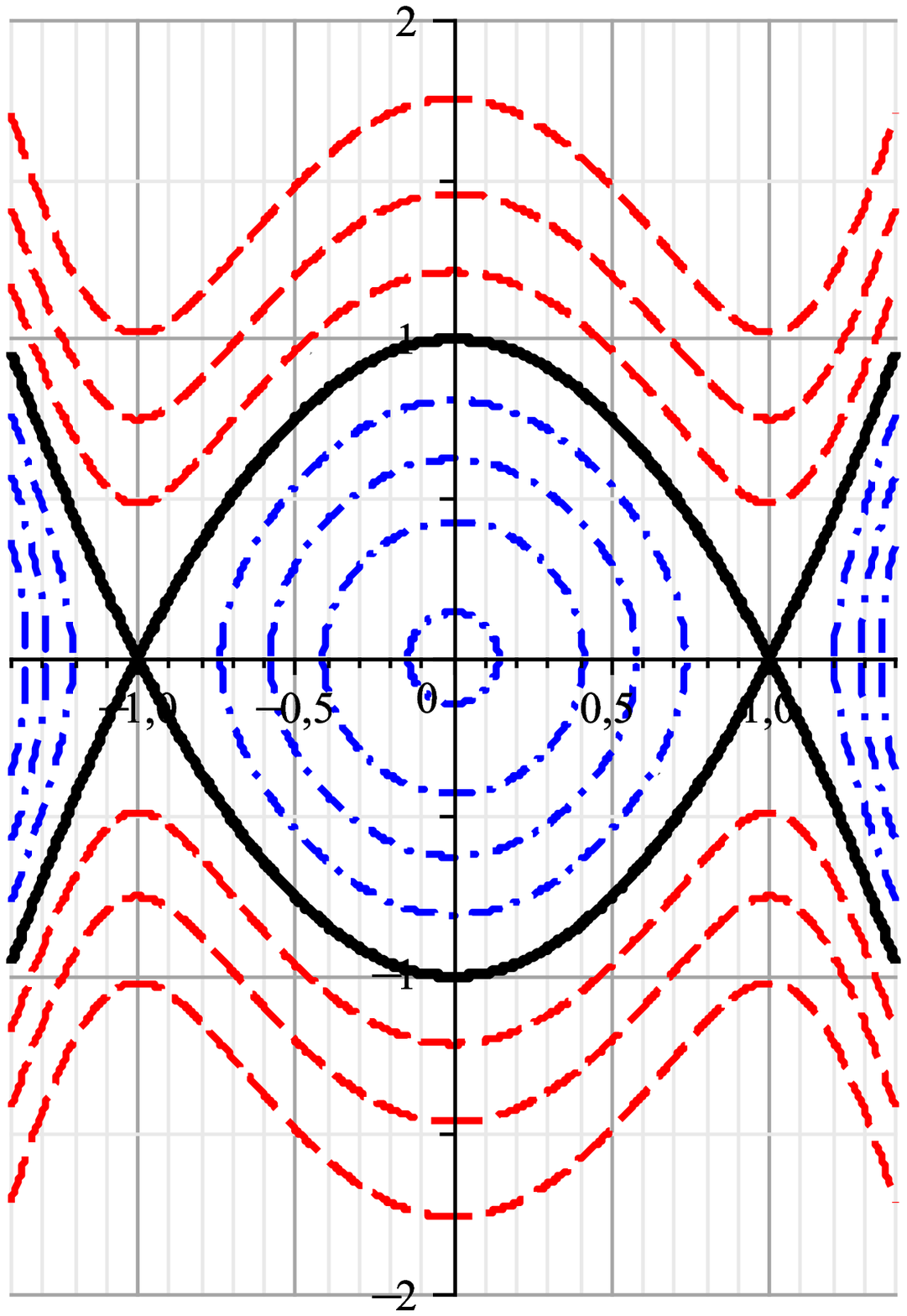}
\vspace{0.08cm}
\includegraphics[width=3.2cm]{st}\hspace{0.8cm}
\includegraphics[width=3.2cm]{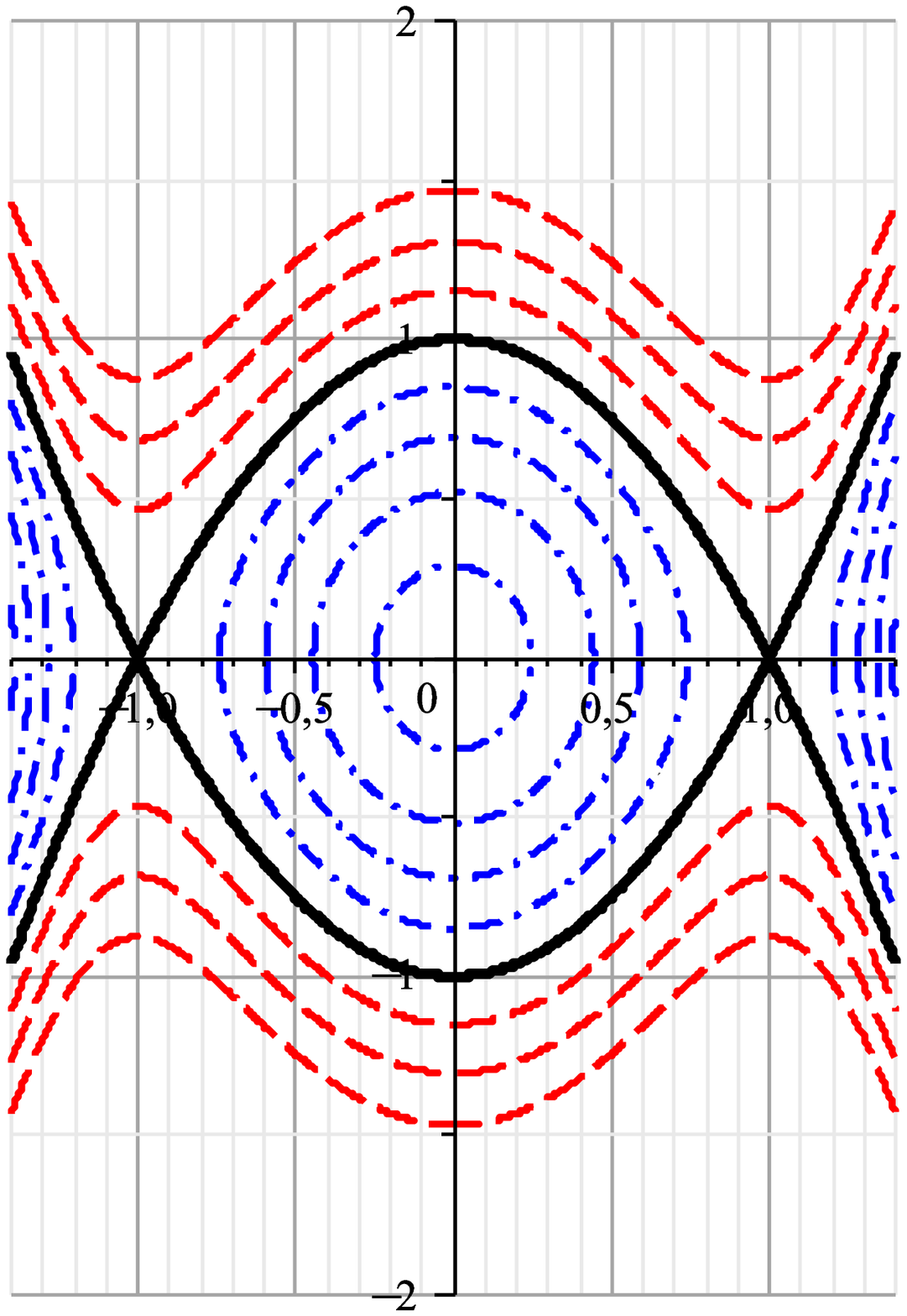}
\vspace{0.07cm}
\includegraphics[width=3.2cm]{st}\hspace{0.8cm}
\includegraphics[width=3.2cm]{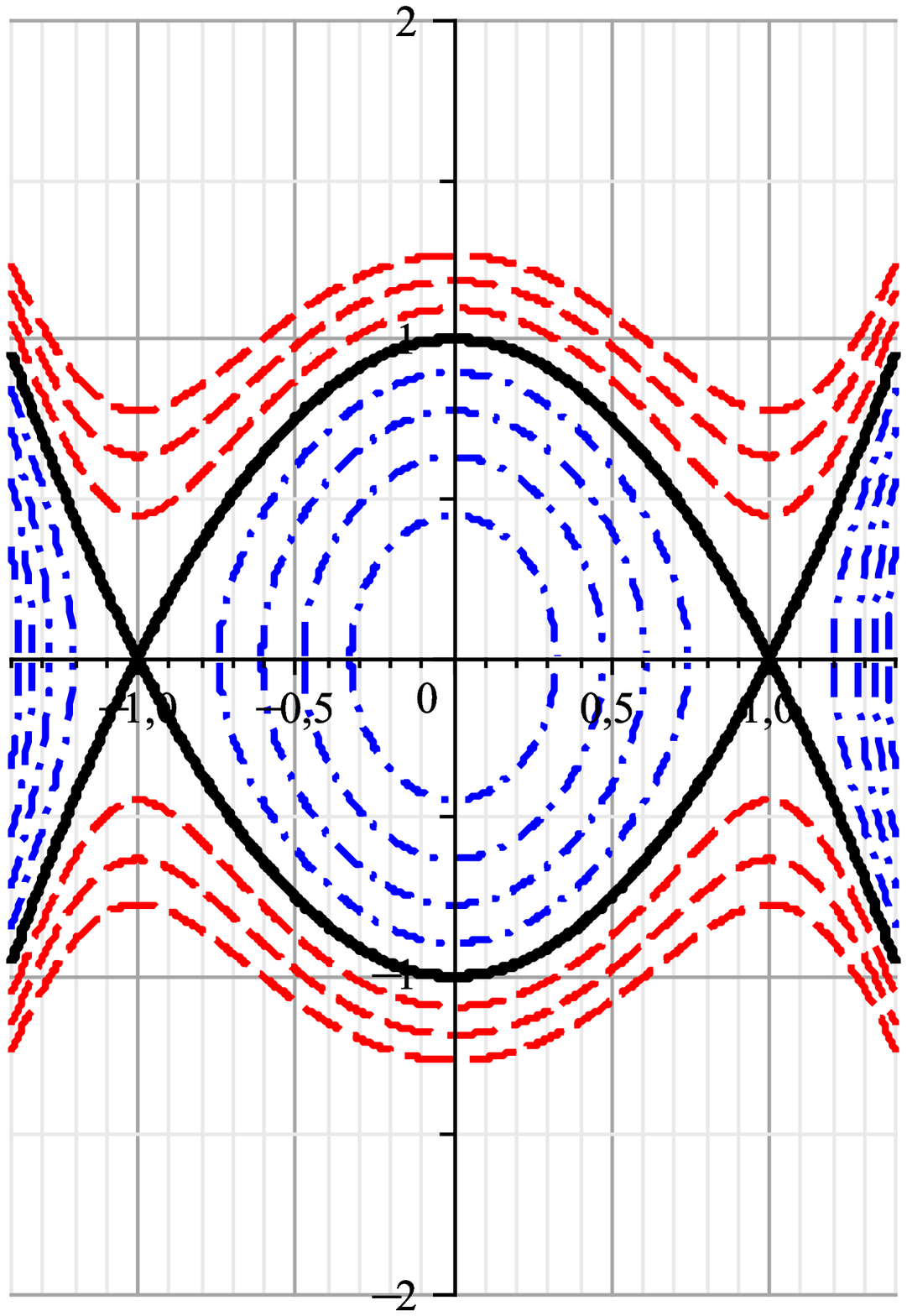}
\caption{Plots of the solutions $\phi(x)$ of the standard (left panel) and ALTW  (right panel) models, for $1/M^{2}=1, 0.5, 0.1,$ from the upper panel to the lower one, respectively. The solid (black) lines stand for $T^{11}=0$, the dashed (red) lines for $T^{11}>0$, and the dot-dashed (blue) lines for $T^{11}<0.$}
\end{figure}

We can investigate another model, with $W_\phi=\cos(\phi)$. In the case of the standard Lagrange density, this corresponds to the sine-Gordon model. Here the equations of motion become
\ben	
\phi^{\prime2}&=&\cos^2(\phi) + 2C,\\
\phi^{\prime2}&=&-1 +\frac{1+\frac{\cos^2(\phi)}{M^2}}{(1-\frac{C}{M^2})^2}.
\een
These equations are investigated in Fig.~2, and there we show how the solutions behave for two distinct values of $1/M^2$. We note that, in terms of $1/M^2$, the models behave in a way similar to the previous case.

We can also investigate models that support nontopological configurations. In this case we follow \cite{non} and we choose
\be
W_\phi=\phi\sqrt{1-\phi\,}.
\ee
The standard model is known as the $\phi^3$ model. Here we get to the equations
\ben	
\phi^{\prime2}&=&\phi^2 (1-\phi) + 2C,\\
\phi^{\prime2}&=&-1 +\frac{1+\frac{\phi^2(1-\phi)}{M^2}}{(1-\frac{C}{M^2})^2}.
\een
These equations are investigated in Fig.~3, and there we see how the two models behave for a single value of $M^2$. In terms of $1/M^2$, the two models also behave in a way similar to the two previous cases.
\begin{figure}[t]
\centering
\includegraphics[width=4.0cm]{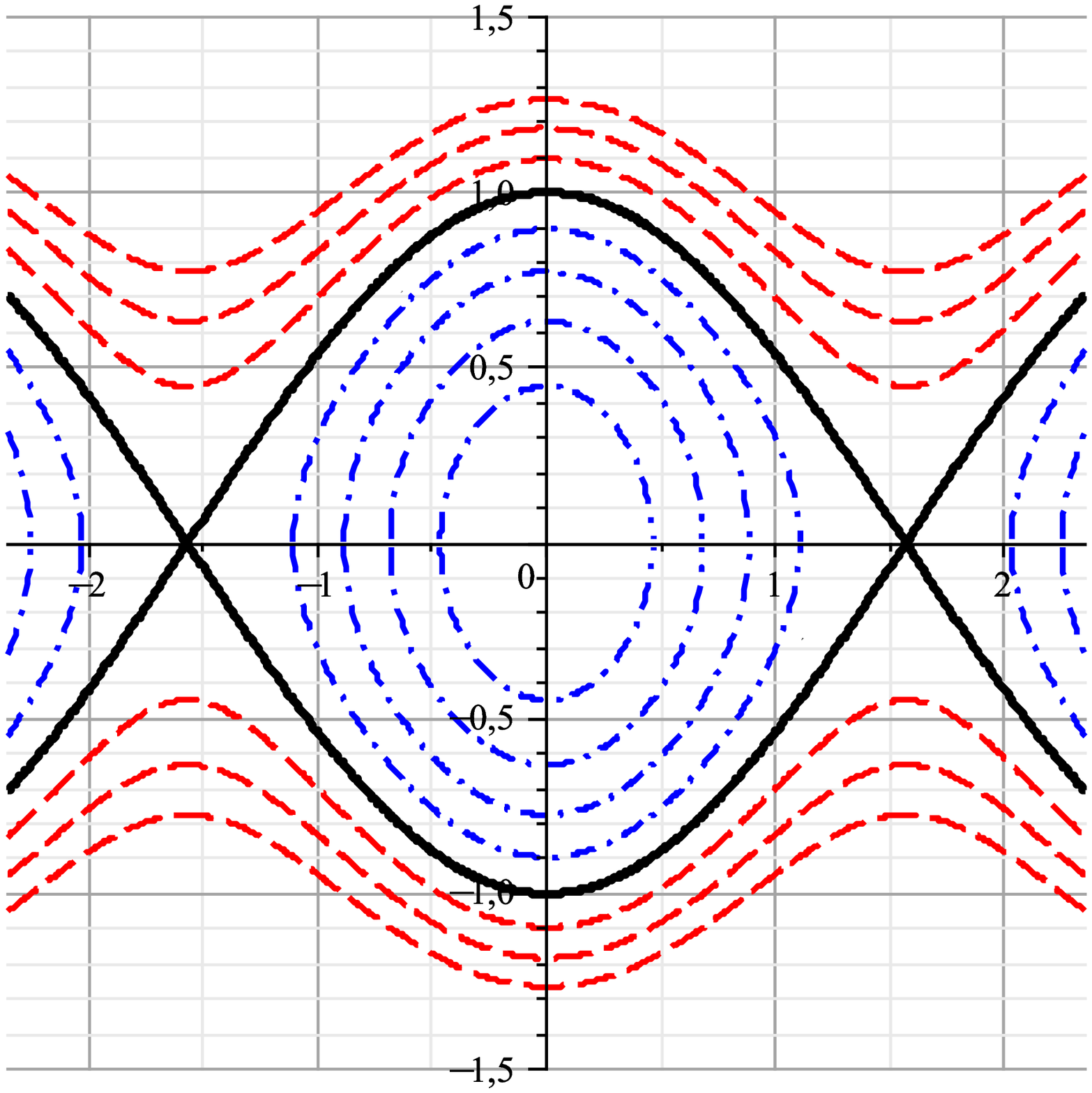}\hspace{0.3cm}
\includegraphics[width=4.0cm]{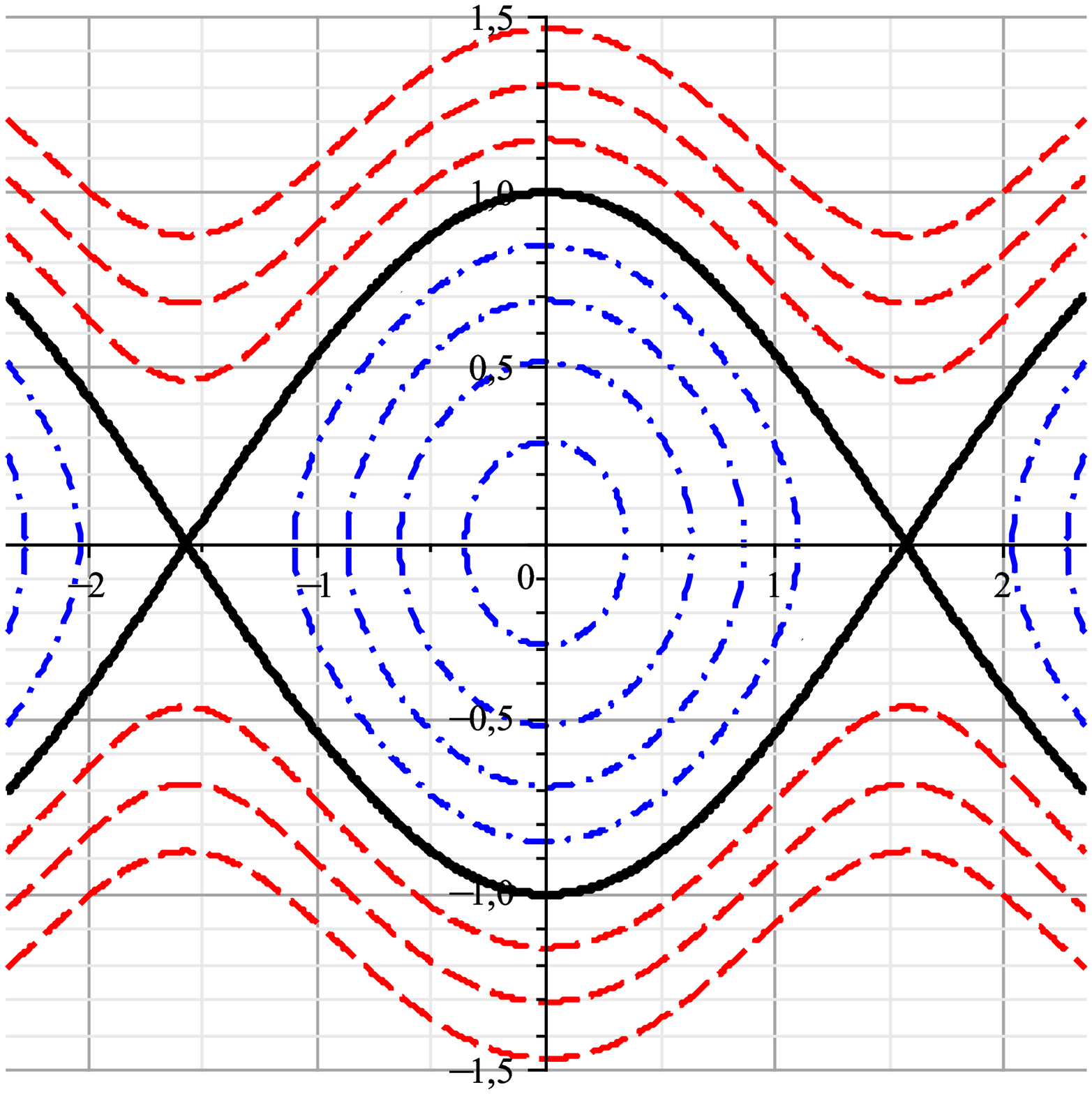}
\vspace{0.7cm}
\includegraphics[width=4.0cm]{SG0}\hspace{0.3cm}
\includegraphics[width=4.0cm]{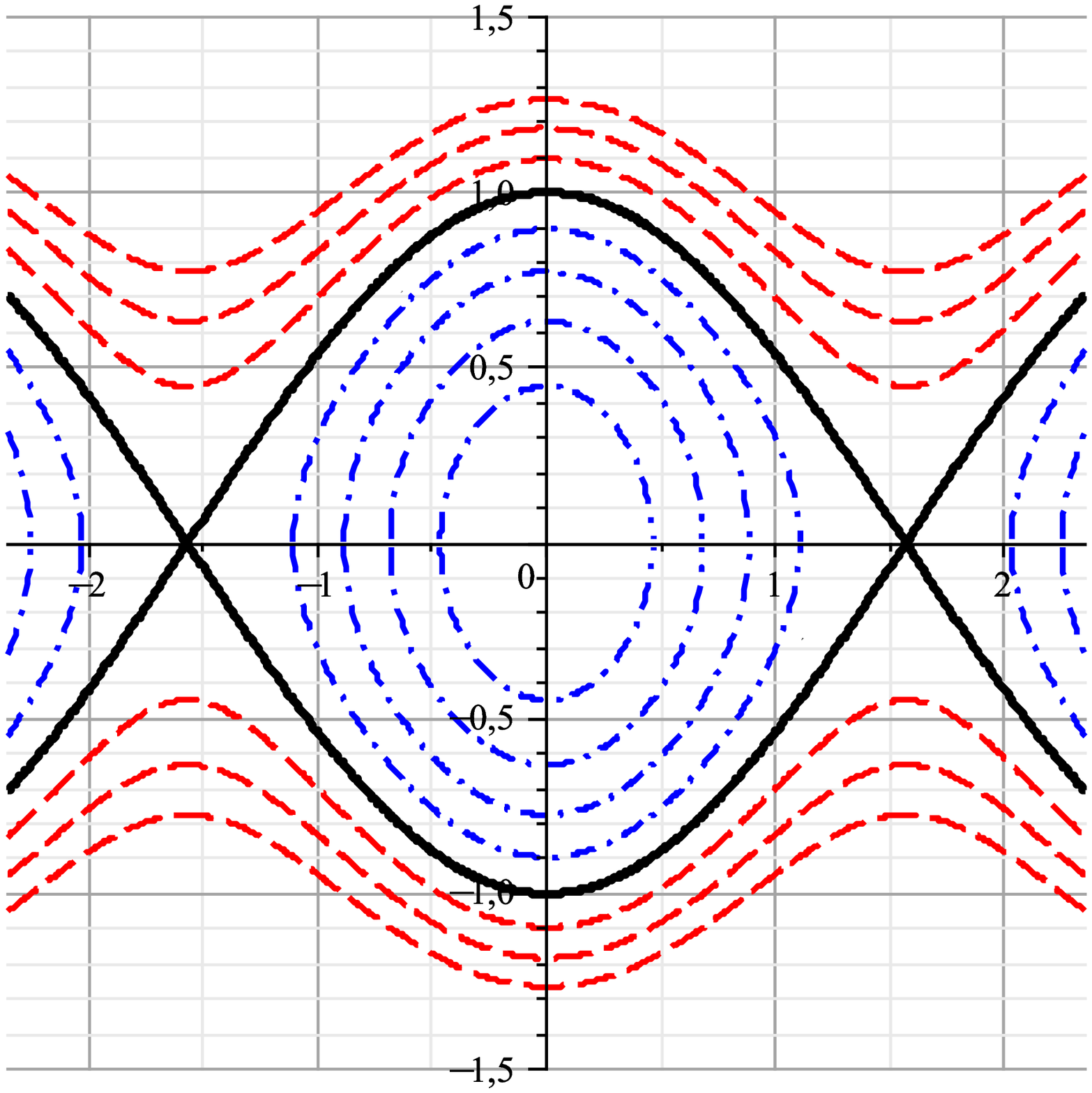}
\caption{Plots of the solutions of the sine-Gordon $V_{sG}(\phi)$ (left panel) and modified (right panel) models, for $1/M^{2}=0.5$ and $0.1$; conventions as in Fig.~1.}
\end{figure}

\begin{figure}[ht]
\centering
\includegraphics[width=4.0cm]{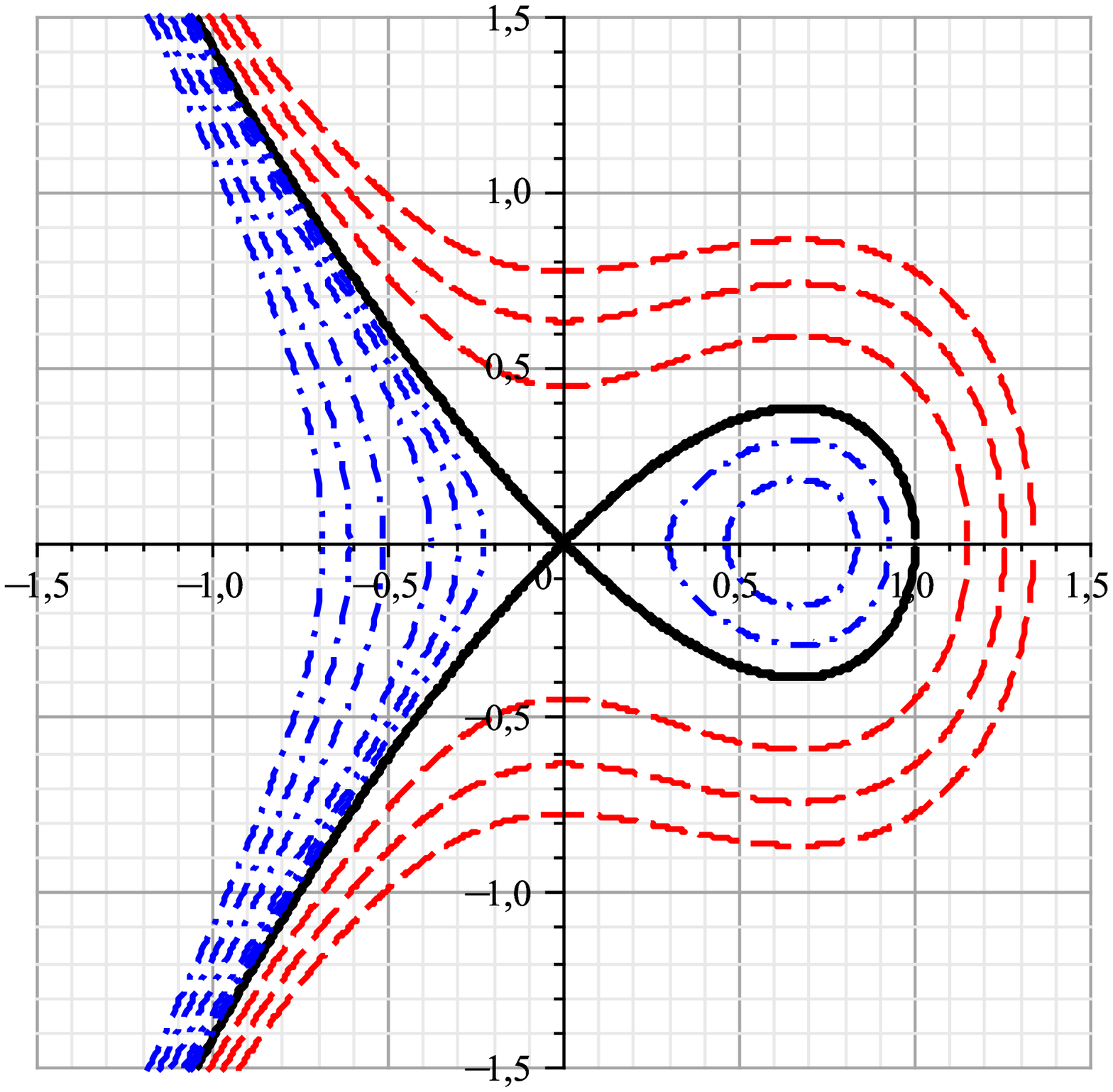}\hspace{0.5cm}
\includegraphics[width=4.0cm]{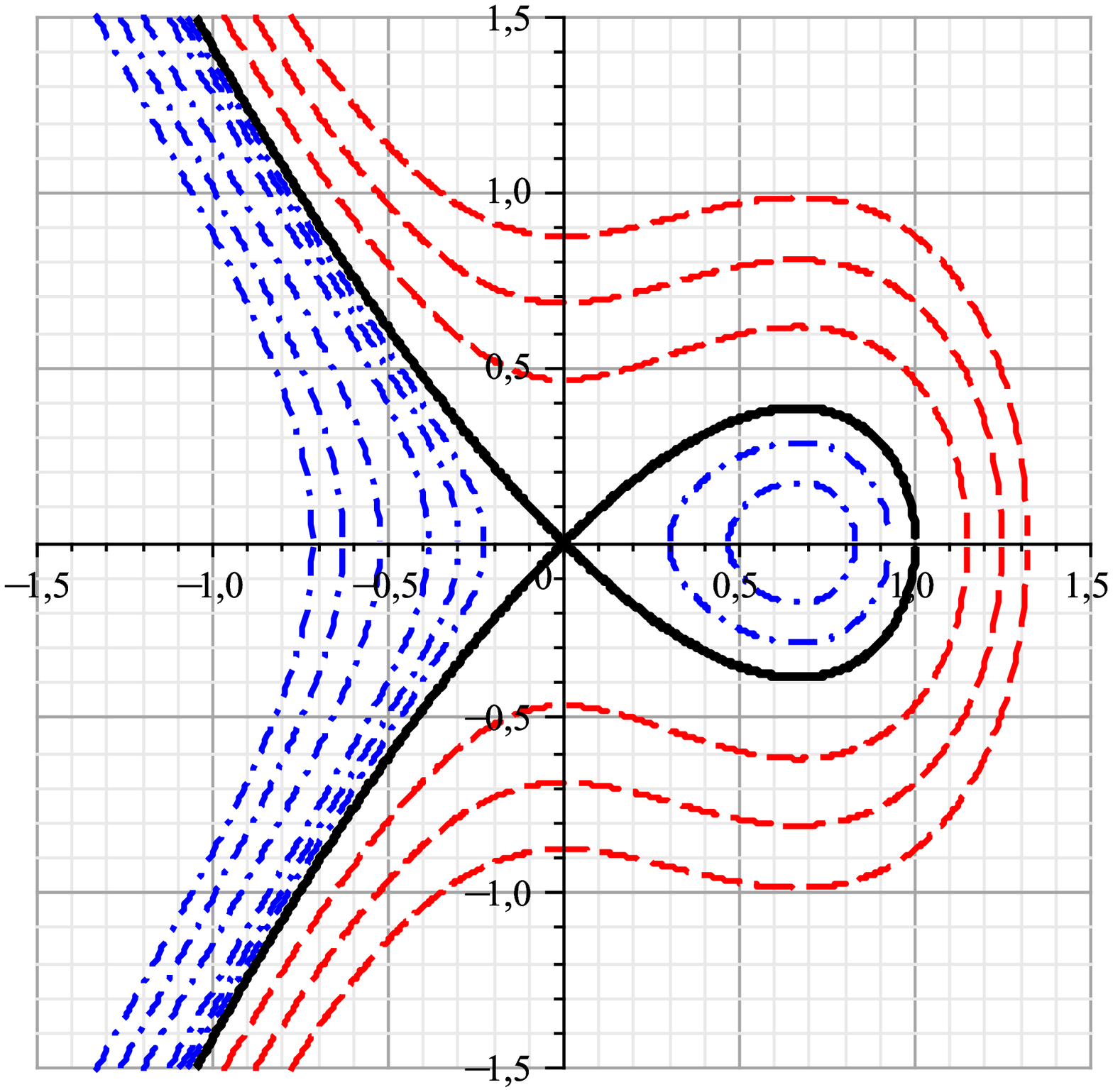}
\caption{Plots of the solutions of the standard $V_3(\phi)$ (left panel) and modified (right panel) models, for $1/M^{2}=0.5$; conventions as in Fig.~1.}
\end{figure}
\subsection{Linear Stability}

Another way to distinguish the twin models is investigating linear stability, as done in Ref.~ \cite{altw}. Here, however, we follow another route. We introduce general fluctuations for the scalar field, $\phi(x,t)=\phi(x)+\eta(x)\cos(\omega\,t)$, where $\phi(x)$ represents the static solution. For a general Lagrange density, we use the fluctuations in the equation of motion to obtain up to first order in $\eta$,
\be
-\left[\left(2\LL_{XX}X+\LL_X\right)\eta^\prime\right]^{\prime}=(\LL_{\phi\phi}+(\LL_{\phi X} \phi^\prime)^\prime+\omega^2 \LL_X)\eta.
\ee

We first suppose that the solution is constant and represents a minimum $v$ of the potential, that is, ${\phi}=v$. It obeys $W_{\phi}(v)=0$, and so we get
\be
-\eta^{\prime\prime}+W^2_{{\phi\phi}}\big|_{v}\eta =\omega^2 \eta,
\ee
which is valid for both the standard and modified models. Therefore, linear stability at the minimum state $v$ does not distinguish the two models.

On the other hand, for static solution $\phi=\phi(x)$, it is possible to distinguish the two twin theories.  In the standard theory, we have the following equation for the fluctuation 
\be\label{Sc}
-\eta^{\prime\prime} +V_q(x)\; \eta = \omega^2 \eta,
\ee
where
\be
V_q(x) ={W_{\phi\phi}^2+W_\phi W_{\phi\phi\phi}},
\ee
is the potential of the Schr\"odingerlike equation \eqref{Sc}. However, in the modified ALTW model, the equation which controls stability is much more complicated. Indeed, it has the general form
\ben
-\left(1+\frac{W_\phi^2}{M^2}\right)^{-1}\!\!{\eta^{\prime\prime}}-\!\left(\left(1+\frac{W_\phi^2}{M^2}\right)^{-1}\right)^{\!\prime} \eta^\prime+&& \\+\nonumber  \left(W_{\phi\phi}^2+W_{\phi}W_{\phi\phi\phi} +\frac{1}{M^2}\left(\frac{W_{\phi}^2W_{\phi\phi}}{1+\frac{W_\phi^2}{M^2}}\right)^{\!\prime}\,\right)\eta &=&\omega^2\eta.
\een
We see that this last equation is different from the previous one. However, instead of following the investigation done in \cite{altw}, here we introduce 
new variables, following the procedure introduced in Ref.~\cite{b3}. We change $x\to z$ and $\eta\to u$ using the expressions
\bes\label{xz}
\ben
dx &=& \left(1+\frac{W_\phi^2}{M^2}\right)^{-\frac12}dz,\\
\eta&=&\left(1+\frac{W_\phi^2}{M^2}\right)^{\frac14}\,\, u,
\een
\ees
which allow writing the Schor\"odingerlike equation
\be
-u_{zz}+ U_q(z)\;u = \omega^2 u,
\ee
where 
\be
U_q(z)=\frac{(\sqrt{\LL A})_{zz}}{\sqrt{\LL A}}- \frac{1}{\LL_X}\left(\LL_{\phi\phi}+\frac{1}{A}\left(\LL_{X\phi}\frac{\phi_z}{A}\right)_{\!\!z}\right).
\ee
It is hard to write an analytical expression $x=x(z)$, for $M$ generic. However, we can consider the case where $M^2$ is very large, in order to obtain analytical results in a perturbative way. In this case we can write, up to the order $1/M^2$, 
\be
x=z-\frac{W(\phi(z))}{M^2},
\ee
with the choice $W(\phi(0))=0$. Moreover, we note that a generic function of the scalar field $\phi$, $F(\phi(x))$, can be expressed in the form, up to the order $1/M^2$,
\be
F(\phi(x))=F(\phi(z))-\frac{(F_\phi W_\phi W)(z)}{2M^2},
\ee
and so the potential can be written in the form
\ben
 U_q(z)&=&U_q^s(z)-\frac{W_\phi}{2M^2}\bigl[3W W_{\phi\phi\phi}+
 W W_\phi W_{\phi\phi\phi\phi}\nonumber\\
 &&+3W_\phi^2 W_{\phi\phi\phi}+8W_\phi W^2_{\phi\phi}\bigr],
\een
where $U_q^s(z)$ stands for the the potential of the standard model.
In the case of $W_{\phi}=1-\phi^2,$ we get
\ben
 U_q(z)&\!=\!&4-6 \,{\rm sech}^2(z)\nonumber\\
 &&\!\!+\frac1{M^2}[4\, {\rm sech}^2(z)\!\!+\!\!14\, {\rm sech}^4(z)\!\!-\!\!21\, {\rm sech}^6(z)].\;\;\;\;\;\;\;
 \een
\begin{figure}[h!]
\centering
\includegraphics[width=0.36\textwidth]{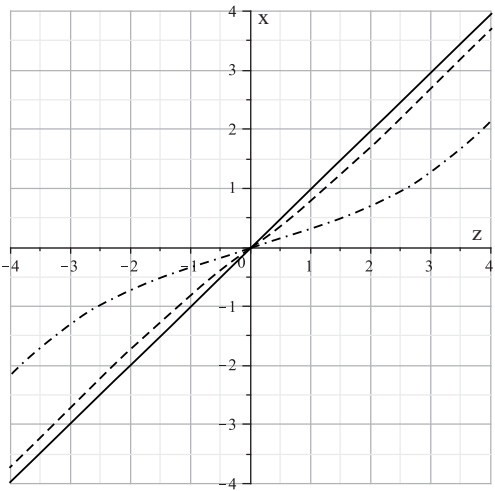}
\caption{Plots of $x=x(z)$ for solid, dashed and dot-dashed lines, for $1/M^{2}=0.1, 1, 10$, respectively.}\label{figuraxz}
\end{figure}

In the general case, we have to rely on numerical integration to obtain $x=x(z)$. In Fig.~\ref{figuraxz} we depict the behavior of $x=x(z)$ for some values of $1/M^2$, taking $W_\phi=1-\phi^2$ in \eqref{xz}.  Note that in the region out of the core of the defect, $\phi$ approaches the minima and $W_\phi$ is small, and we can approximate $z\approx x + c$ where $c$ is  a constant, and $u\approx \eta$.

\begin{figure}[h!]
\centering
\includegraphics[width=0.36\textwidth]{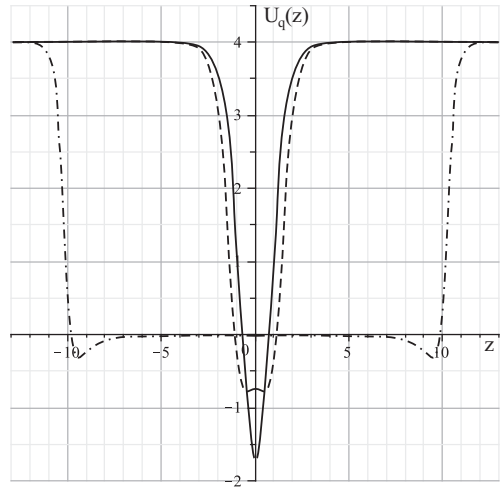}
\caption{Plots of the potential $U_q(z)$ for solid, dashed and dot-dashed lines, for $1/M^{2}=0.1, 1, 10$, respectively.\label{pPOT}}
\end{figure}

\begin{figure}[h!]
\centering
\includegraphics[height=5.7cm]{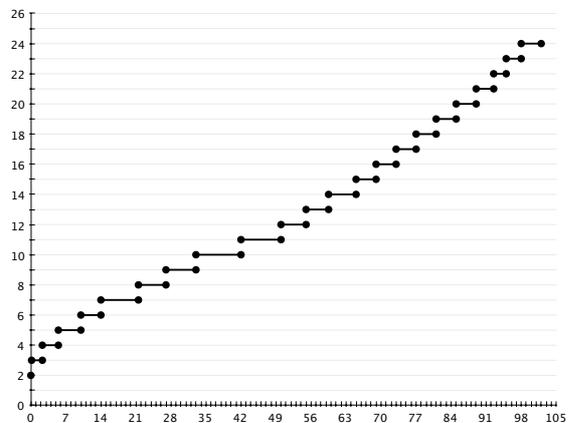}
\caption{Number of bound states supported by the potential $U_q(z)$ as a function of $1/M^2$.\label{figuraDOT}}
\end{figure}

In the general case, the full expression of the potential is given by 
\ben
U_q(z)&=& U_q^s (x(z)) - \frac{\frac{W_{\phi}^2}{M^2}}{\left(1+\frac{W_\phi^2}{M^2}\right)^3} (x(z))\nonumber\\ 
 &\times&\Biggl[W_\phi W_{\phi\phi\phi} \left(\frac32 + \frac52\frac{W_\phi^2}{M^2}+\left(\frac{W_\phi^2}{M^2}\right)^2\right)\nonumber\\ 
 &+& W_{\phi\phi}^2 \left(\!4\!+\!\frac{13}4 \frac{W_{\phi}^2}{M^2} \!+\! \left(\frac{W_\phi^2}{M^2}\right)^2\right) \Biggr](x(z)).
 \een
We use this expression to depict the potential in Fig.~{\ref{pPOT}} in the case of $W_\phi=1-\phi^2$, for some values of $1/M^2$. In the figure, we see that the potential starts as the modified P\"osch-Teller potential, and approximates the case of a box, for increasing $1/M^2$. We note that although the depth of the box does not depend on $M$, its width increases as $1/M^2$ increases, adding more and more bound states into the box. Similar result was also obtained in Ref.~{\cite{altw}}. In the general case, we can numerically count the number of bound states as $1/M^2$ increases. The result is depicted in Fig.~\ref{figuraDOT}, where the dots indicate the value of $1/M^2$
for which the number of bound states jumps to the next one.


\section{Other Twinlike Models}
\label{sec4}

Let us now focus on the issue of obtaining other twinlike models. Toward this goal, we consider a more general Lagrange density, given by \cite{s3}
\be\label{newmodel}
{\cal L}=\dfrac{2^{n-1}}{n}X|X|^{n-1}-V(\phi).
\ee
We note that the case $n=1$ leads us back to the standard model. For other values of $n$, we are changing kinematics, and this is of interest in cosmology \cite{c1,c2,c3,c4}
and also in the context of compactons, that is, of defect structures which lives in a compact region within the real line \cite{s1,s2,s3}. We use the more general model 
\eqref{newmodel} to write 
\be
{\cal L}_X=2^{n-1} |X|^{n-1}.
\ee
We now follow the procedure introduced in \cite{b3} and for static solution we get
\bes\label{ori}
\ben
\phi'&=&W_{\phi}^{\frac{1}{2n-1}},
\label{plf}
\\
V(\phi)&=&\dfrac{2n-1}{2n}W_{\phi}^{\frac{2n}{2n-1}},
\een\ees
with the energy density given by
\be
\rho(x)=W_{\phi}^{\frac{2n}{2n-1}}.
\label{t00f}
\ee

We suppose that the twin theory has the following form
\be
{\cal L}=M^2+M^2F(X)G(\phi),
\ee
where $M$ is the mass parameter, as before, and $F(X)$ and $G(\phi)$ are functions to be determined.  For static solutions, we can write
\bes\label{mod}
\ben
M^2 F_X\,G(\phi)\phi^{\prime}&=&W_{\phi},
\label{fx} \\
G(\phi)&=&\frac{1}{2XF_X-F}.
\label{ufx}
\een
\ees
Using both \eqref{ori} and \eqref{mod}, and considering that the solution $\phi(x)$ and $W$ are the same for both theories, 
we get to the following differential equation 
\be
F_X\left(M^2-2^nX|X|^{n-1}\right)=-F|2X|^{n-1},
\ee
which can be integrated to give
\be
F(X)=\left(M^2-2^nX |X|^{n-1}\right)^{\frac{1}{2n}}.
\ee
Therefore, the twinlike theory is given by
\be\label{newtwin}
{\cal L}=M^2+M^{\frac{2n+1}{n}}G(\phi)\left(1-2^n \frac{X|X|^{n-1}}{M^2}\right)^{\frac{1}{2n}}.
\ee
The function $G(\phi)$ in (\ref{ufx}) can be written as
\ben
G(\phi)=-M^{\frac{-1}{n}}\left(1+\!\frac{W_{\phi}^{\frac{2n}{2n-1}}}{M^2}\right)^{\frac{2n-1}{2n}},
\een
and the Lagrange density gets to the form 
\be\label{newtwin2}
{\cal L}=M^2\!-\!M^2\left(\!1\!+\!\frac{2n}{2n\!-\!1}\frac{V(\phi)}{M^2}\right)^{\!\!\frac{2n-1}{2n}}\!\!\!\!\left(\!1\!-\!2^n \frac{X|X|^{n-1}}{M^2}\right)^{\!\frac{1}{2n}}\!{.}
\ee
We can use \eqref{upot} to rewrite the above result in terms of $U(\phi)$. It has the form
\ben
{\cal L}=M^2\!\!&-&\!\!M^2\left[1-\frac{n}{2n-1}\left(1-\left(1+\frac{U(\phi)}{M^2}\right)^2\right)\right]^{\!\!\frac{2n-1}{2n}}\nonumber\\
&\times&\left(1-2^n \frac{X|X|^{n-1}}{M^2}\right)^{\!\frac{1}{2n}}\!{.}
\een
Here we note that the above expression exactly reproduces the ALTW model described by \eqref{altw}, in the limit $n\to 1$.
We also note that the energy density is also given by (\ref{t00f}), so it behaves as in the standard model, as expected.

We can investigate the case in which $1/M^2$ is very small. Here we get, up to the order $1/M^2$, 
\ben
{\cal L}&\approx &\dfrac{2^{n-1}}{n}X|X|^{n-1}-V\nonumber\\
&&+ \frac{1}{M^2} \frac{2n\!-\!1}{8n^2}\! \left(\!2^n X |X|^{n-1}\!+\!\frac{2n}{2n-1}V(\phi)\!\right)^2{\! .}
\een
Thus, the twinlike model \eqref{newtwin2} reproduces the model \eqref{newmodel} in the limit $1/M^2\to0$. This is similar to the previous case, so we can say that the two models
\eqref{newmodel} and \eqref{newtwin2} are twinlike models, and they are generalizations of the previous models, described by \eqref{standard} and \eqref{altw}.
It is interesting to remark that the above result was obtained within the first-order framework put forward in Ref.~\cite{b3}.

\section{The Braneworld Scenario}
\label{sec5}

The defect structure which appears in scalar field theory can be used to mimic thick brane in the braneworld scenario with warped geometry and a single extra dimension of infinite extent \cite{RS,GW,F}. In this sense, we can ask how the twinlike models studied before can be used to mimic similar scenario within the braneworld context, in the form of a twinlike configuration. 

To investigate this issue, we follow \cite{b4} and we consider a five-dimensional action describing gravity coupled to the scalar field in the form
\be
S=\int d^5x\sqrt{g}\left(-\dfrac{1}{4}R+{\cal L}\left(\phi,X\right)\right),
\ee
where we are using $4\pi G=1$ and
\be
X=\dfrac{1}{2}\nabla_M\phi\nabla^M\phi,
\ee
with $M,N=0,1,2,3,4$, running on the five-dimensional spacetime. The equation of motion which we obtain for the scalar field in this new scenario is given by
\be\label{beqs}
G^{AB}\nabla_A\nabla_B\phi+2X{\cal L}_{X\phi}-{\cal L}_{\phi}=0,
\ee
where $G^{AB}$ has the form
\be
G^{AB}={\cal L}_Xg^{AB}+{\cal L}_{XX}\nabla^A\phi\nabla^B\phi.
\ee
The energy-momentum tensor becomes
\be
T_{AB}=\nabla_A\phi\nabla_B\phi{\cal L}_X-g_{AB}{\cal L}.
\ee
The line element of the five-dimensional spacetime can be written as $ds^2=e^{2A}\eta_{\mu\nu}dx^{\mu}dx^{\nu}-dy^2$,
where $A$ is used to describe the warp factor. We suppose that both $A$ and $\phi$ are static, such that they only depend on the extra dimension $y$, that is, $A=A(y)$ and $\phi=\phi(y)$. In this case, the equation of motion \eqref{beqs} reduces to
\be
\left({\cal L}_X+2X{\cal L}_{XX}\right)\phi''-\left(2X{\cal L}_{X\phi}-{\cal L}_{\phi}\right)=-4{\cal L}_X\phi'A',
\label{m}
\ee
where prime now denotes derivative with respect to the extra dimension. Also, we are using that 
${\cal L}_X=\partial {\cal L}/\partial X$ and ${\cal L}_{\phi}=\partial {\cal L}/\partial \phi$, etc.

We now focus attention on Einstein equations, which lead us to the equations
\bes
\ben
A''&=&\frac{4}{3}X{\cal L}_X,
\label{a}\\
A'^2&=&\frac{1}{3}\left({\cal L}-2X{\cal L}_X\right),
\label{b}
\een
\ees
where $X=-\phi'^2/2$ for static configuration, as before. If we use (\ref{m}), we can write
\be
\left({\cal L}-2X{\cal L}_X\right)'=-4\phi'^2A'{\cal L}_X.
\ee
To get to the first-order framework, we suppose that
\be
A'=-\frac{1}{3}W(\phi),
\label{al}
\ee
In this case, the equations \eqref{a} and \eqref{b} leads us to, respectively,
\bes
\ben\label{c}
{\cal L}_X\phi'&=&\frac{1}{2}W_{\phi}, \\
{\cal L}-2X{\cal L}_X&=&\frac{1}{3}W^2.
\label{d}
\een
\ees

In the case of a standard scalar field, the Lagrange density has the form \eqref{standard}, and so the equation of motion (\ref{m}) becomes
\be
\phi''+4A'\phi'=V_{\phi}.
\label{pll}
\ee
After substituting this result in (\ref{c}) and (\ref{d}), we get to the following first-order equation
\be
\phi'=\frac{1}{2}W_{\phi},
\label{pl}
\ee
and the potential
\be
V=\frac{1}{8}W_{\phi}^2-\frac{1}{3}W^2.
\label{v}
\ee
Note that the equations (\ref{al}) and (\ref{pl}) solve equation (\ref{pll}) for the potential (\ref{v}). Note also that the energy density is given by
\be
T^{00}=e^{2A}\left(\frac{1}{4}W_{\phi}^2-\frac{1}{3}W^2\right).
\label{t00}
\ee

This is the standard situation, but now we have to find the twin model. Toward this goal, 
let us consider a scalar field theory governed by the following Lagrange density
\be\label{mbrane}
{\cal L}=M^2-M^2\left(1+\frac{U}{M^2}\right)\sqrt{1-\frac{2X}{M^2}}+f(\phi),
\ee
where $f(\phi)$ is to be determined. We use the equations (\ref{c}) and (\ref{d}) to write, respectively
\be
\frac{1+\frac{U}{M^2}}{\sqrt{1-\frac{2X}{M^2}}}\phi'=\frac{1}{2}W_{\phi},
\label{pl2}
\ee
\be
\frac{1+\frac{U}{M^2}}{\sqrt{1-\frac{2X}{M^2}}}=1,
\label{u}
\ee
where we have used $f(\phi)=W^2/3$, such that, if we substitute \eqref{u} into \eqref{pl2} we get to
\be
\phi^\prime=\frac12 W_\phi,
\ee
and therefore we get the same equation of motion of the standard model. Moreover, we can use \eqref{u} to write
\be
\phi'^2=2U+ \frac{U^2}{M^2}.
\ee

Therefore, the Lagrange density of the twin brane model has the following form
\be\label{mbrane1}
{\cal L}=M^2-M^2\left(1+\frac{U}{M^2}\right)\sqrt{1-\frac{2X}{M^2}}+\frac{1}{3}W^2.
\ee
Moreover, the energy density is
\be
T^{00}=e^{2A}\left(\frac{1}{4}W_{\phi}^2-\frac{1}{3}W^2\right),
\ee
which exactly reproduces the previous expression (\ref{t00}). Thus, the two models have the same solution, with the very same energy density. This is a feature of twinlike models, as expected. In this case, the potential can be written as
\be
U(\phi)=-M^2+\frac{1}{2}M^2\sqrt{4+\frac{W_{\phi}^2}{M^2}}.
\ee
We use this to write
\be\label{ubrane}
U(\phi)=-M^2+M^2\sqrt{1+2\frac{V(\phi)}{M^2}+\frac{2}{3}\frac{W^2(\phi)}{M^2}},
\ee
which relates the potentials of the two theories. Here we note that, in the limit of very small $1/M^2$, we can use the above expression \eqref{ubrane} to verify that the twinlike model \eqref{mbrane1} lead us back to the standard model, as expected. 

\section{Comments and Conclusions}
\label{sec6}

In this work we extended the idea put forward in \cite{altw}, concerning the presence of two distinct scalar field theories which support the same defect structure. We called the two field theory models twinlike models, since they engender the same defect structure, with the very same energy density. 

We have shown that the two models can be distinguished classically, using the fact that $T^{11}$ is constant, but it does not need to be zero. For nonzero $T^{11}$, the two equations of motion give distinct static solutions. Distinct results also appear when one investigates linear stability of the defect structure in the two models. This was firstly presented in \cite{altw}, but here we have investigated stability in the standard way, searching for the Schr\"odingerlike equation which usually appears in the investigation of linear stability. Since the ALTW model is described by a modified Lagrange density, of the k-defect type, it does not lead us directly to a Schr\"odinger equation. However, we have found the correct way to obtain the Schr\"odinger equation, and this allowed us to search for the number of bound states connected to the defect.  

We have extended the idea of obtaining twinlike models to a new class of field theory models. This is interesting since it shows that the presence of twinlike models is not specific of the pair of models presented in \cite{altw}. In this sense, the presence of twinlike models seems to deserve further investigation, since it may be a generic situation in Field Theory. 
 
Another line of investigation concerns the presence of twinlike models within the braneworld context \cite{RS}. As one knows, scalar fields can be used to generate thick branes \cite{GW,F}, and so we have extended the idea of twinlike models to the braneworld scenario. The investigation led us to the twinlike braneworld concept, in which a brane can be generated from two distinct models of real scalar fields, coupled to gravity in $(4,1)$ dimensions, with a single extra dimension of spacelike nature and infinite extent. 

The existence of twinlike models poses an interesting issue, concerning the coupling of scalar fields with fermions, which is usually of the Yukawa type. In this case, one knows that fermions may fractionalize in the presence of topological defect \cite{JR}, like the ones we have studied in this work. In this sense, although one can construct two distinct scalar field models to couple to the fermion field, if the two scalar field models are twinlike models, the defect structure they may engender will induce the very same quantum corrections to the fermions, at least at the lowest quantum level. Thus, in the presence of fermions, it seems that we need to go beyond the first quantum corrections, in order for the fermions to distinguish the two  models. Another issue refers to the coupling with gauge fields, to study the presence of twinlike models for vortices and monopoles. These and other related issues are presently under consideration, and we hope to report on them in the near future.

We would like to thank CAPES and CNPq, Brazil, and FCT project CERN/FP/116358/2010, Portugal, for partial financial support.

\end{document}